# The generalization of A. E. Kennelly theory of complex representation of the electrical quantities in sinusoidal periodic regime to the one and three-phase electric quantities in non-sinusoidal periodic regime

Gheorghe Mihai

University of Craiova, Faculty of Electrotechnics, Bvd Decebal 107, 200440, Romania

E-mail: gmihai@elth.ucv.ro

Abstract. In this paper, a new mathematical method of electrical circuits calculus is proposed based on the theory of the complex linear operators in matrix form. The newly proposed method generalizes the theory of complex representation of electrical quantities in sinusoidal periodic regime to the non-sinusoidal periodic regime.

Contents

_______________________________________________________________

# 1. Introduction

The method of the complex representation of electrical quantities is applicable only in the case of sinusoidal periodic regime and not in the case of non sinusoidal periodic regime [4]. In order to also be used in non-sinusoidal periodic regime, it is first used the superposition principle through which, each harmonica takes action in the electrical circuit. The superposition principle for the electric currents can be applied only after one has returned in real time domain. The theory of the complex linear operators allows considering the well-known relations from the theory of complex representation of the sinusoidal electrical quantities, only with a different mathematical structure.

## 2. One-phased electric circuits in non-sinusoidal periodic regime

2.1. Electric quantities represented by complex linear operators
2.1.1. The operational form for the voltage, electric current and impedance

Let us consider a side of an electric circuit with R, L, C parameters. The voltage and electric current on that side of the circuit are as follows:

$$u(t) = U_0 + \sum_{k=1}^{\infty} U_k \sin(k\omega t + \alpha_k) \tag{1}$$

and

$$i(t) = I_0 + \sum_{k=1}^{\infty} \sqrt{2} I_k \sin(k\omega t + \alpha_k + \varphi_k) \tag{2}$$

Accordingly to each harmonica from relations (1) and (2), we have the complex representation:

$$u_k(t) \Rightarrow \underline{U}_k = U_k (\cos\alpha_k + j\sin\alpha_k) = U_k^{'} + jU_k^{''} \qquad k = 1,2,....\infty \tag{3}$$

$$i_k(t) \Rightarrow \underline{I}_k = I_k (\cos(\alpha_k + \varphi_k) + j\sin(\alpha_k + \varphi_k)) = I_k^{'} + jI_k^{''} \qquad k = 1,2,3....\infty \tag{4}$$

The complex impedance for that side of circuit, with R, L, C in series, is:

$$\underline{Z}_k = R + j\left(k\omega L - \frac{1}{k\omega C}\right) \tag{5}$$

It is known that for each harmonica, k, we can write:

$$\underline{U}_k = \underline{Z}_k \underline{I}_k \qquad k = 1,2,3....\infty \tag{6}$$

Accordingly to the proposed method, we create a diagonal matrix, infinitely extended with elements from relations (3) and (4):

$$\hat{\underline{U}} = \begin{pmatrix} U_0 & 0 & 0 & 0 \\ 0 & \underline{U}_1^{'} + j\underline{U}_1^{''} & 0 & 0 \\ 0 & 0 & ..... & 0 \\ 0 & 0 & 0 & \underline{U}_k^{'} + j\underline{U}_k^{''} \end{pmatrix}$$

$$\tag{7}$$

$$\hat{\underline{I}} = \begin{pmatrix} I_0 & 0 & 0 & 0 \\ 0 & \underline{I}_1' + j\underline{I}_1'' & 0 & 0 \\ 0 & 0 & ..... & 0 \\ 0 & 0 & 0 & \underline{I}_k' + j\underline{I}_k'' \end{pmatrix}$$

In order to respect the superposition principle imposed by relation (6), it results that the impedance matrix has to be a diagonal one:

$$\hat{\underline{Z}} = \begin{pmatrix} \underline{Z}_0 & 0 & 0 & 0......... \\ 0 & \underline{Z}_1 & 0 & 0......... \\ 0 & 0 & \underline{Z}_2 & 0......... \\ 0 & 0 & .......... & \underline{Z}_k \end{pmatrix} \qquad (8)$$

From relations (7) and (8), it results that relation (6) has the following generalized form:

$$\hat{\underline{U}} = \hat{\underline{Z}} \cdot \hat{\underline{I}} \qquad (9)$$

2.1.2. The operational form for the electric resistance, inductive reactance and capacitive reactance

We will proceed in detailing relation (8) in more elements with matrix form, such as:
- electric resistance:

$$\hat{R} = \begin{pmatrix} R & ... & ... & ... \\ ... & R & ... & ... \\ ... & ... & R & ... \\ ... & ... & ... & R \end{pmatrix} \qquad (10)$$

- inductive reactance:

$$j\hat{X}_L = j\begin{pmatrix} 0\omega L & ... & ... & ... \\ ... & 1\omega L & ... & ... \\ ... & ... & 2\omega L & ... \\ ... & ... & ... & k\omega L \end{pmatrix} = j\omega L \begin{pmatrix} 0 & ... & ... & ... \\ ... & 1 & ... & ... \\ ... & ... & 2 & ... \\ ... & ... & ... & k \end{pmatrix} \qquad (11)$$

We introduce the harmonics operator $\hat{K}$, as follows:

$$\hat{K} = \begin{pmatrix} 0 & ... & ... & ... \\ ... & 1 & ... & ... \\ ... & ... & 2 & ... \\ ... & ... & ... & k \end{pmatrix} \quad (12)$$

Relation (11) can be written using the harmonics operator $\hat{K}$ :

$$j\hat{X}_L = j\omega L\hat{K} \quad (13)$$

which generalizes relation: $jX_L = jk\omega L$

- capacitive reactance:
-

$$-j\hat{X}_C = j\begin{pmatrix} \dfrac{1}{0\omega C} & ... & ... & ... \\ ... & \dfrac{1}{1\omega C} & ... & ... \\ ... & ... & ... & ... \\ ... & ... & ... & \dfrac{1}{k\omega C} \end{pmatrix} = -j\dfrac{1}{\omega C}\begin{pmatrix} \dfrac{1}{0} & ... & ... & ... \\ ... & \dfrac{1}{1} & ... & ... \\ ... & ... & ... & ... \\ ... & ... & ... & \dfrac{1}{k} \end{pmatrix} \quad (14)$$

Relation (14) can be written using the harmonics operator $\hat{K}$ :

$$-j\hat{X}_C = -j\dfrac{1}{\omega C}\hat{K}^{-1} = \left(j\omega C\hat{K}\right)^{-1} \quad (15)$$

which generalizes relation:

$$-jX_C = \dfrac{1}{jk\omega C}$$

## 2.1.3. Effective value of the electric voltage and current

The effective value for the electric current and voltage is obtained from relations (1) and (2):

$$\begin{cases} U_{ef} = \sqrt{U_0^2 + U_1^2 + .... + U_k^2 + ......} = \sqrt{\displaystyle\sum_{k=0}^{\infty} U_k} \\[3mm] I_{ef} = \sqrt{I_0^2 + I_1^2 + .... + I_k^2 + ......} = \sqrt{\displaystyle\sum_{k=0}^{\infty} I_k^2} \end{cases} \quad (16)$$

Based on relations (7), it can be noticed that relations (16) are obtained from:

$$\left|\underline{\hat{U}}\right| = \sqrt{trace\,\underline{\hat{U}}\,\underline{\hat{U}}^{*}} \qquad\qquad \left|\underline{\hat{I}}\right| = \sqrt{trace\,\underline{\hat{I}}\,\underline{\hat{I}}^{*}} \qquad (17)$$

The quantities $\underline{\hat{U}}^{*}$ and $\underline{\hat{I}}^{*}$, represent the complex conjugated form of the matrixes (7).

### 2.1.4. Formulas for returning in real time domain

Knowing the matrix form of the electric current $\underline{\hat{I}}$, we want to determine a way to return in real time domain, meaning to get back to relation (2).

In sinusoidal periodic regime, if one knows the complex form for the electric current $\underline{I}$, in order to return in real time domain it uses the formulas:

$$\begin{cases} \varphi = arctg\,\dfrac{\mathbf{Im}\,ag\cdot\underline{I}}{\mathbf{Re}\,al\cdot\underline{I}} \\[2mm] \left|\underline{I}\right| = \sqrt{\left(\mathbf{Re}\,al\,\underline{I}\right)^{2} + \left(\mathbf{Im}\,ag\,\underline{I}\right)^{2}} \\[2mm] i(t) = \sqrt{2}\left|\underline{I}\right|\sin\left(\omega t - \varphi\right) \end{cases} \qquad (18)$$

Accordingly to the proposed method, the returning in real time is made symbolically with the same sets of formulas (18), to which is introduced the symbol" ^". For (Σ) we introduce: trace.

Based on the above mentioned, formulas (18) take the following operational form:

$$\begin{cases} \hat{\varphi} = arctg\left[\left(\mathbf{Im}\,ag\cdot\underline{\hat{I}}\right)\left(\mathbf{Re}\,al\cdot\underline{\hat{I}}\right)^{-1}\right] \\[2mm] \left|\underline{\hat{I}}\right| = \sqrt{\left(\mathbf{Re}\,al\,\underline{\hat{I}}\right)^{2} + \left(\mathbf{Im}\,ag\,\hat{I}\right)^{2}} \\[2mm] i(t) = trace\left[\sqrt{2}\left|\underline{\hat{I}}\right|\sin\left(\hat{k}\omega t - \hat{\varphi}\right)\right] \end{cases} \qquad (19)$$

Accordingly to relation (7) we have:

$$\begin{cases} \mathbf{Re}\,al\,\underline{\hat{I}} = \dfrac{1}{2}\left(\underline{\hat{I}} + \underline{\hat{I}}^{*}\right) = \begin{pmatrix} I_{0} & \cdots & \cdots & \cdots \\ \cdots & I_{1}^{'} & \cdots & \cdots \\ \cdots & \cdots & \cdots & \cdots \\ \cdots & \cdots & \cdots & I_{k}^{'} \end{pmatrix} \end{cases} \qquad (20)$$

$$\mathbf{Im}\,ag\,\underline{\hat{I}} = \frac{-j}{2}\left(\underline{\hat{I}} - \underline{\hat{I}}^*\right) = \begin{pmatrix} 0 & ... & ... & ... \\ ... & I_1^{''} & ... & ... \\ ... & ... & ... & ... \\ ... & ... & ... & I_k^{''} \end{pmatrix}$$

From relations (19) and (20) we obtain:

$$\hat{\varphi} = arctg \begin{pmatrix} 0 & ... & ... & ... \\ ... & \dfrac{I_1^{''}}{I_1^{'}} & ... & ... \\ ... & ... & ... & ... \\ ... & ... & ... & \dfrac{I_k^{''}}{I_k^{'}} \end{pmatrix} \qquad (21)$$

Due to the fact that the matrix (21) is diagonal, the trigonometry function is applied to each element of the matrix:

$$\hat{\varphi} = \begin{pmatrix} 0 & ... & I_1^{''} & ... & ... \\ ... & arctg\,\dfrac{I_1^{''}}{I_1^{'}} & ... & ... \\ ... & ... & ... & ... \\ ... & ... & ... & arctg\,\dfrac{I_k^{''}}{I_k^{'}} \end{pmatrix} \qquad (22)$$

From relations (19) and (20) we obtain:

$$\left|\underline{\hat{I}}\right| = \sqrt{\left(\mathbf{Re}\,al\,\underline{\hat{I}}\right)^2 + \left(\mathbf{Im}\,ag\,\underline{\hat{I}}\right)^2} = \begin{pmatrix} I_0 & ... & ... & ... \\ ... & \sqrt{I_1^{'2} + I_1^{''2}} & ... & ... \\ ... & ... & ... & ... \\ ... & ... & ... & \sqrt{I_k^{'2} + I_k^{''2}} \end{pmatrix} \qquad (23)$$

Accordingly to the symbols from relation (19) and relations (12) and (22), we have:

$$\sin\left(\omega t \hat{k} - \hat{\varphi}\right) = \sin\left[\left[\left(\omega t \begin{pmatrix} 0 & ... & ... & ... \\ ... & 1 & ... & ... \\ ... & ... & 2 & ... \\ ... & ... & ... & k \end{pmatrix} - \begin{pmatrix} 0 & ... & ... & ... \\ ... & arctg\,\dfrac{I_1^{''}}{I_1^{'}} & ... & ... \\ ... & ... & ... & ... \\ ... & ... & ... & arctg\,\dfrac{I_k^{''}}{I_k^{'}} \end{pmatrix}\right)\right]\right] =$$

$$= \begin{pmatrix} 0 & \cdots & \cdots & \cdots \\ \cdots & \sin\left(\omega t - arctg\,\dfrac{I_1^{''}}{I_1^{'}}\right) & \cdots & \cdots \\ \cdots & \cdots & \cdots & \cdots \\ \cdots & \cdots & \cdots & \sin\left(k\omega t - arctg\,\dfrac{I_k^{''}}{I_k^{'}}\right) \end{pmatrix} \quad (24)$$

The electric current's expression in real time domain, on each harmonica, is obtained from relations (23) and (24) and is given by relation:

$$\hat{I}(t) = \begin{pmatrix} 0 & \cdots & \cdots & \cdots \\ \cdots & I_1\sin(\omega t - \varphi_1) & \cdots & \cdots \\ \cdots & \cdots & \cdots & \cdots \\ \cdots & \cdots & \cdots & I_k\sin(k\omega t - \varphi_k) \end{pmatrix} \quad (25)$$

From relation (25), the Fourier series is obtained:

$$i(t) = I_0 + trace\sqrt{2}\hat{I}(t) \quad (26)$$

### 2.1.5. The operational form of the electric powers

The complex electric power in sinusoidal periodic regime is calculated using the relation:

$$\underline{S} = \underline{U} \cdot \underline{I}^{*} \quad (27)$$

Accordingly to the proposed method, in non-sinusoidal periodic regime, relation (27) can be rewritten in a symbolic form:

$$\underline{\hat{S}} = \underline{\hat{U}} \cdot \underline{\hat{I}}^{*} \quad (28)$$

which based on relations (7), has the expression:

$$\underline{\hat{S}} = \begin{pmatrix} \underline{U}_1\underline{I}_1^{*} & \cdots & \cdots & \cdots \\ \cdots & \underline{U}_2\underline{I}_2^{*} & \cdots & \cdots \\ \cdots & \cdots & \cdots & \cdots \\ \cdots & \cdots & \cdots & \underline{U}_k\underline{I}_k^{*} \end{pmatrix} \quad (29)$$

The active and reactive power can be obtained from (29) and has the expression:

$$
\begin{cases}
\hat{P} = \mathbf{Re}\,al\,\underline{\hat{S}} = \frac{1}{2}\left(\underline{\hat{S}} + \underline{\hat{S}}^*\right) = \begin{pmatrix} P_1 & \dots & \dots & \dots \\ \dots & P_2 & \dots & \dots \\ \dots & \dots & \dots & \dots \\ \dots & \dots & \dots & P_k \end{pmatrix} \\[4em]
\hat{Q} = \mathbf{Im}\,ag\,\underline{\hat{S}} = \frac{-j}{2}\left(\underline{\hat{S}} - \underline{\hat{S}}^*\right)\begin{pmatrix} Q_1 & \dots & \dots & \dots \\ \dots & Q_2 & \dots & \dots \\ \dots & \dots & \dots & \dots \\ \dots & \dots & \dots & Q_k \end{pmatrix}
\end{cases}
\tag{30}
$$

The active electric power, respectively the reactive power on all harmonics is:

$$
P_{all} = trace\hat{P} \qquad\qquad Q_{all} = trace\hat{Q} \tag{31}
$$

The apparent power in form of:

$$
S^2 = \left(U_0^2 + U_1^2 + \dots\dots + U_k^2 + \dots\right)\cdot\left(I_0^2 + I_1^2 + \dots\dots + I_k^2 + \dots\right) \tag{32}
$$

is obtained from relations (16):

$$
S^2 = \left(trace\underline{\hat{U}}\,\underline{\hat{U}}^*\right)\left(trace\underline{\hat{I}}\,\underline{\hat{I}}^*\right) \tag{33}
$$

The deforming power as Budeanu defines [3] is obtained from relations (31) and (33):

$$
D^2 = S^2 - P_{all}^2 - Q_{all}^2 \tag{34}
$$

## 3. Three-phased electric circuits in non-sinusoidal periodic regime

### 3.1. The operational form of the three-phased electric voltages and currents

Let us consider a three-phased electric circuit, with electromotive voltages as follows:

$$
u(t)_{ei} = u_{ei0} + \sum_{k=1}^{\infty} \sqrt{2}u_{eik}\sin\left(k\omega t + \alpha_{ik}\right) \quad i = 1,2,3 \tag{35}
$$

In the three-phased electric circuit, the electrical currents are determined:

$$
i_i(t) = I_{i0} + \sum_{k=1}^{\infty} \sqrt{2}I_{ik}\sin\left(k\omega t + \alpha_{ik} + \varphi_{ik}\right) \quad i = 1,2,3 \tag{36}
$$

The phases' electric parameters are: $(R_1, L_1, C_1)$, $(R_2, L_2, C_2)$, $(R_3, L_3, C_3)$.

Accordingly to relation (7), relations (35) and (36) can be written in operational form:

$$\hat{U}_{ei} = \begin{pmatrix} U_{ei0} & \cdots & \cdots & \cdots \\ \cdots & U_{ei1} & \cdots & \cdots \\ \cdots & \cdots & \cdots & \cdots \\ \cdots & \cdots & \cdots & U_{eik} \end{pmatrix} \qquad i = 1,2,3 \qquad (37)$$

$$\hat{I}_i = \begin{pmatrix} \underline{I}_{i0} & \cdots & \cdots & \cdots \\ \cdots & \underline{I}_{i1} & \cdots & \cdots \\ \cdots & \cdots & \cdots & \cdots \\ \cdots & \cdots & \cdots & \underline{I}_{ik} \end{pmatrix} \qquad i = 1,2,3 \qquad (38)$$

In sinusoidal periodic regime, the line voltages are defined as follows:

$$\underline{U}_{ij} = \underline{U}_i - \underline{U}_j \qquad i = 1,2,3 , \ j = 1,2,3 , \ i \neq j \qquad (39)$$

In non-sinusoidal periodic regime, relation (39) is rewritten as a relation between the operators:

$$\hat{\underline{U}}_{ij} = \hat{\underline{U}}_i - \hat{\underline{U}}_j \qquad i = 1,2,3 , \ j = 1,2,3 , \ i \neq j \qquad (40)$$

The following operators correspond to phase impedances:

$$\hat{\underline{Z}}_i = \begin{pmatrix} \underline{Z}_{i0} & \cdots & \cdots & \cdots \\ \cdots & \underline{Z}_{i1} & \cdots & \cdots \\ \cdots & \cdots & \cdots & \cdots \\ \cdots & \cdots & \cdots & \underline{Z}_{ik} \end{pmatrix} \qquad i = 1,2,3 \qquad (41)$$

## 3.2. Solving methods for the electric circuits in non-sinusoidal periodic regime

### 3.2.1. The neutral point displacement voltage

Among the specific methods for calculating the three-phased electric circuits in star connection, one can use the neutral point displacement voltage.

In sinusoidal permanent regime the neutral point displacement voltage is expressed by relation [3]:

$$\Delta \underline{U} = -(\underline{Y}_1 + \underline{Y}_2 + \underline{Y}_3 + \underline{Y}_4) \cdot (\underline{U}_{e1} \underline{Y}_1 + \underline{U}_{e2} \underline{Y}_2 + \underline{U}_{e3} \underline{Y}_3) \qquad (42)$$

which, for non-sinusoidal periodic regime, becomes:

$$\varDelta \hat{\underline{U}} = -\left(\hat{\underline{Y}}_1 + \hat{\underline{Y}}_2 + \hat{\underline{Y}}_3 + \hat{\underline{Y}}_4\right)^{-1} \cdot \left(\hat{\underline{U}}_{e1}\hat{\underline{Y}}_1 + \hat{\underline{U}}_{e2}\underline{Y}_2 + \hat{\underline{U}}_{e3}\hat{\underline{Y}}_3\right) \tag{43}$$

where:

$$\hat{\underline{Y}}_i = \begin{pmatrix} \dfrac{1}{\underline{Z}_{i0}} & \cdots & \cdots & \cdots \\ \cdots & \dfrac{1}{\underline{Z}_{i1}} & \cdots & \cdots \\ \cdots & \cdots & \cdots & \cdots \\ \cdots & \cdots & \cdots & \dfrac{1}{\underline{Z}_{ik}} \end{pmatrix} \qquad i = 1,2,3 \tag{44}$$

The phase voltages for the consumer become:

$$\hat{\underline{U}}_i = \hat{\underline{U}}_{ei} + \varDelta \hat{\underline{U}} \qquad i = 1,2,3 \tag{45}$$

Knowing the operational form of the consumer's voltage and its operational impedance, the electric currents can be calculated using the formulas:

$$\hat{\underline{I}}_i = \left(\hat{\underline{Z}}_i\right)^{-1} \hat{\underline{U}}_i \qquad i = 1,2,3 \tag{46}$$

## 3. 2. 2. The method of the symmetric components

In non-sinusoidal periodic regime, to operators $1, a, a^2 \left(a = -\dfrac{1}{2} + j\dfrac{\sqrt{3}}{2}\right)$ correspond the operators:

$$\hat{1} = \begin{pmatrix} 1 & \cdots & \cdots & \cdots \\ \cdots & 1 & \cdots & \cdots \\ \cdots & \cdots & 1 & \cdots \\ \cdots & \cdots & \cdots & 1 \end{pmatrix} \tag{47}$$

$$\hat{a} = \begin{pmatrix} a & \cdots & \cdots & \cdots \\ \cdots & a & \cdots & \cdots \\ \cdots & \cdots & a & \cdots \\ \cdots & \cdots & \cdots & a \end{pmatrix} \tag{48}$$

$$\hat{a}^2 = \begin{pmatrix} \hat{a}^2 & ... & ... & ... \\ ... & \hat{a}^2 & ... & ... \\ ... & ... & \hat{a}^2 & ... \\ ... & ... & ... & \hat{a}^2 \end{pmatrix} \qquad (49)$$

The calculus of the symmetric components which correspond to a given system of nonsymmetrical three-phased quantities, in non-sinusoidal periodic regime is using the following operational relations:

$$\begin{cases} \underline{\hat{U}}_h = \dfrac{1}{3}\left(\underline{\hat{U}}_1 + \underline{\hat{U}}_2 + \underline{\hat{U}}_3\right) \\ \underline{\hat{U}}_d = \dfrac{1}{3}\left(\underline{\hat{U}}_1 + \hat{a}\underline{\hat{U}}_2 + \hat{a}^2\underline{\hat{U}}_3\right) \\ \underline{\hat{U}}_i = \dfrac{1}{3}\left(\underline{\hat{U}}_1 + \hat{a}^2\underline{\hat{U}}_2 + \hat{a}\underline{\hat{U}}_3\right) \end{cases} \qquad (50)$$

Accordingly to operators (47), (48), (49), we have the impedances:

$$\begin{cases} \underline{\hat{Z}}_h = \dfrac{1}{3}\left(\underline{\hat{Z}}_1 + \underline{\hat{Z}}_2 + \underline{\hat{Z}}_3\right) \\ \underline{\hat{Z}}_d = \dfrac{1}{3}\left(\underline{\hat{Z}}_1 + \hat{a}\underline{\hat{Z}}_2 + \hat{a}^2\underline{\hat{Z}}_3\right) \\ \underline{\hat{Z}}_i = \dfrac{1}{3}\left(\underline{\hat{Z}}_1 + \hat{a}^2\underline{\hat{Z}}_2 + \hat{a}\underline{\hat{Z}}_3\right) \end{cases} \qquad (51)$$

Between the symmetrical components of the voltage system (50) and the ones of the current system there are [3]:

$$\begin{pmatrix} \underline{\hat{Z}}_h & \underline{\hat{Z}}_d & \underline{\hat{Z}}_i \\ \underline{\hat{Z}}_i & \underline{\hat{Z}}_h & \underline{\hat{Z}}_d \\ \underline{\hat{Z}}_d & \underline{\hat{Z}}_i & \underline{\hat{Z}}_h \end{pmatrix} \begin{pmatrix} \hat{I}_h \\ \hat{I}_d \\ \hat{I}_i \end{pmatrix} = \begin{pmatrix} \underline{\hat{U}}_h \\ \underline{\hat{U}}_d \\ \underline{\hat{U}}_i \end{pmatrix} \qquad (52)$$

The system of equations (52), is solved using the symbolic expressions of the operators $\underline{\hat{I}}_h, \underline{\hat{I}}_d$ and $\underline{\hat{I}}_i$.

$$\begin{pmatrix} \hat{\underline{I}}_h \\ \hat{\underline{I}}_d \\ \hat{\underline{I}}_i \end{pmatrix} = \begin{pmatrix} \underline{\hat{Z}}_h & \underline{\hat{Z}}_d & \underline{\hat{Z}}_i \\ \underline{\hat{Z}}_i & \underline{\hat{Z}}_h & \underline{\hat{Z}}_d \\ \underline{\hat{Z}}_d & \underline{\hat{Z}}_i & \underline{\hat{Z}}_h \end{pmatrix}^{-1} \begin{pmatrix} \underline{\hat{U}}_h \\ \underline{\hat{U}}_d \\ \underline{\hat{U}}_i \end{pmatrix} \qquad (53)$$

After calculus we obtain:

$$\begin{cases} \underline{\hat{I}}_h = \left( \underline{\hat{Z}}_h^3 + \underline{\hat{Z}}_d^3 + \underline{\hat{Z}}_i^3 - 3\underline{\hat{Z}}_h \cdot \underline{\hat{Z}}_d \cdot \underline{\hat{Z}}_i \right)^{-1} \left[ \left( \underline{\hat{Z}}_h^2 - \underline{\hat{Z}}_d \underline{\hat{Z}}_i \right) \underline{\hat{U}}_h + \left( \underline{\hat{Z}}_i^2 - \underline{\hat{Z}}_h \underline{\hat{Z}}_d \right) \underline{\hat{U}}_d + \left( \underline{\hat{Z}}_d^2 - \underline{\hat{Z}}_i \underline{\hat{Z}}_h \right) \underline{\hat{U}}_i \right] \\[2mm] \underline{\hat{I}}_d = \left( \underline{\hat{Z}}_h^3 + \underline{\hat{Z}}_d^3 + \underline{\hat{Z}}_i^3 - 3\underline{\hat{Z}}_h \cdot \underline{\hat{Z}}_d \cdot \underline{\hat{Z}}_i \right)^{-1} \left[ \left( \underline{\hat{Z}}_d^2 - \underline{\hat{Z}}_i \underline{\hat{Z}}_h \right) \underline{\hat{U}}_h + \left( \underline{\hat{Z}}_h^2 - \underline{\hat{Z}}_d \underline{\hat{Z}}_i \right) \underline{\hat{U}}_d + \left( \underline{\hat{Z}}_i^2 - \underline{\hat{Z}}_h \underline{\hat{Z}}_d \right) \underline{\hat{U}}_i \right] \\[2mm] \underline{\hat{I}}_i = \left( \underline{\hat{Z}}_h^3 + \underline{\hat{Z}}_d^3 + \underline{\hat{Z}}_i^3 - 3\underline{\hat{Z}}_h \cdot \underline{\hat{Z}}_d \cdot \underline{\hat{Z}}_i \right)^{-1} \left[ \left( \underline{\hat{Z}}_i^2 - \underline{\hat{Z}}_h \underline{\hat{Z}}_d \right) \underline{\hat{U}}_h + \left( \underline{\hat{Z}}_d^2 - \underline{\hat{Z}}_i \underline{\hat{Z}}_h \right) \underline{\hat{U}}_d + \left( \underline{\hat{Z}}_h^2 - \underline{\hat{Z}}_d \underline{\hat{Z}}_i \right) \underline{\hat{U}}_i \right] \end{cases}$$

$$(54)$$

In relations (54) all terms from right are diagonal square matrixes with known values, so that $\underline{\hat{I}}_h, \underline{\hat{I}}_d, \underline{\hat{I}}_i$ are diagonal matrixes with known values. The electric currents written using operational form on each phase become known and are calculated with formulas:

$$\begin{cases} \underline{\hat{I}}_1 = \underline{\hat{I}}_h + \underline{\hat{I}}_d + \underline{\hat{I}}_i \\ \underline{\hat{I}}_2 = \underline{\hat{I}}_h + \hat{a}^2 \underline{\hat{I}}_d + \hat{a} \underline{\hat{I}}_i \\ \underline{\hat{I}}_3 = \underline{\hat{I}}_h + \hat{a} \underline{\hat{I}}_d + \hat{a}^2 \underline{\hat{I}}_i \end{cases} \qquad (55)$$

Returning in real time domain is made accordingly to the ones presented in paragraph 2.1.4.

## 3.3. Electric power in three-phased electric circuits

### 3. 3. 1. Active and reactive complex electric power

The complex electric power of a phase can be calculated using relations:

$$\underline{\hat{S}}_i = \underline{\hat{U}}_i \cdot \underline{\hat{I}}_i^* = \begin{pmatrix} \underline{U}_{i1} \underline{I}_{i1}^* & \dots & \dots & \dots \\ \dots & \underline{U}_{i2} \underline{I}_{i2}^* & \dots & \dots \\ \dots & \dots & \dots & \dots \\ \dots & \dots & \dots & \underline{U}_{ik} \underline{I}_{ik}^* \end{pmatrix} \qquad i = 1,2,3 \qquad (56)$$

The active and reactive electric powers on each phase depending on the harmonics are:

$$\hat{P}_i = \operatorname{Re} al \underline{\hat{S}}_i = \frac{1}{2} \left( \underline{\hat{S}}_i + \underline{\hat{S}}_i^* \right) = \begin{pmatrix} P_{i1} & \dots & \dots & \dots \\ \dots & P_{i2} & \dots & \dots \\ \dots & \dots & \dots & \dots \\ \dots & \dots & \dots & P_{ik} \end{pmatrix} \qquad i = 1,2,3 \qquad (57)$$

$$\hat{Q}_i = \operatorname{Im} ag \underline{\hat{S}}_i = \frac{-j}{2} \left( \underline{\hat{S}}_i - \underline{\hat{S}}_i^* \right) = \begin{pmatrix} Q_{i1} & \dots & \dots & \dots \\ \dots & Q_{i2} & \dots & \dots \\ \dots & \dots & \dots & \dots \\ \dots & \dots & \dots & Q_{ik} \end{pmatrix} \qquad i = 1,2,3 \qquad (58)$$

The total active and reactive electric powers of the three-phased electric circuit are:

$$P_{all} = \sum_{i=1}^{3} trace\hat{P}_i \qquad (59)$$

$$Q_{all} = \sum_{i=1}^{3} trace\hat{Q}_i \qquad (60)$$

### 3.3.2. The total maximum power of the three-phased electric circuit

Accordingly to relation (32), the total apparent power of the three-phased electric circuit in non-sinusoidal regime has the following expression [5]:

$$S = \left( \sum_{i=1}^{3} \sum_{k=1}^{\infty} U_{ik}^2 \right) \cdot \left( \sum_{i=1}^{3} \sum_{k=1}^{\infty} I_{ik}^2 \right) \qquad (61)$$

Depending on the voltage and current operators, relation (61) becomes:

$$S = \left[ \sum_{i=1}^{3} trace\left( \hat{\underline{U}}_i \hat{\underline{U}}_i^* \right) \right]\left[ \sum_{i=1}^{3} trace\left( \hat{\underline{I}}_i \hat{\underline{I}}_i^* \right) \right] \qquad (62)$$

### 3.3.3. The non-symmetric electric power

The expression for the non-symmetric electric power of the three-phased electric circuit depending on the harmonics is:

$$P_{nosym.all}^2 = \sum_{i=1}^{3} \sum_{\substack{j=1 \\ i \neq j}}^{3} \left[ \sum_{k=1}^{\infty} U_{ik}^2 I_{jk}^2 + U_{jk}^2 I_{ik}^2 - 2U_{ik}U_{jk}I_{ik}I_{jk} \cos\left( \varphi_{ik} - \varphi_{jk} \right) \right] \qquad (63)$$

In order to determine the expression of the non-symmetry power of the three-phased electric circuit corresponding to the harmonic **k**, we need to know the following expressions for the powers:
- the total active power of the three-phased electric circuit corresponding to the harmonic **k**:

$$P_{k,all} = \sum_{i=1}^{3} U_{ik}I_{ik}\cos\varphi_{ik} = \sum_{i=1}^{3} P_{ik}$$

- the total reactive power of the three-phased electric circuit corresponding to the harmonic **k**:

$$Q_{k,all} = \sum_{i=1}^{3} U_{ik}I_{ik}\sin\varphi_{ik} = \sum_{i=1}^{3} Q_{ik}$$

- the total apparent power of the three-phased electric circuit corresponding to the harmonic **k**:

$$S_{k,all} = \sum_{i=1}^{3} U_{ik} I_{ik} = \sum_{i=1}^{3} S_{ik}$$

The non-symmetry power of the three-phased electric circuit corresponding to the harmonic **k** is given by:

$$P_{k,nosym}^2 = S_{k,all}^2 - P_{k,all}^2 - Q_{k,all}^2 \qquad (64)$$

The non-symmetry power of the three-phased electric circuit corresponding to all harmonics is obtained by summing relation (64) after **k**:

$$P_{nosym,all}^2 = \sum_{k=1}^{\infty} P_{k,nosym}^2 = \sum_{k=1}^{\infty} \left( S_{k,all}^2 - P_{k,all}^2 - Q_{k,all}^2 \right) \qquad (65)$$

Depending on relations (56), (57) and (58), relation (65) becomes:

$$P_{nosym,all}^2 = trace\left[ \hat{S}\hat{S}^{\bullet} - \hat{P}_{all}^2 - \hat{Q}_{all}^2 \right] \qquad (66)$$

In which:

$$\begin{cases} \hat{S} = \sum_{i=1}^{3} \hat{S}_i \\ \hat{S}^{\bullet} = \sum_{i=1}^{3} \hat{S}_i^{\bullet} \end{cases} \qquad\qquad \begin{cases} \hat{P}_{all} = \sum_{i=1}^{3} \hat{P}_i \\ \hat{Q}_{all} = \sum_{i=1}^{3} \hat{Q}_i \end{cases}$$

### 3.3.4. The total deforming electric power

The total deforming electric power represents the sum of all deforming powers corresponding to each phase and has the following expression:

$$D_{all}^2 = \sum_{i=1}^{3} \left[ \sum_{\substack{k=1 \\ l \neq k}}^{\infty} \sum_{l=1}^{\infty} U_{ik}^2 I_{il}^2 + U_{il}^2 I_{ik}^2 - 2U_{ik}^2 U_{il}^2 I_{ik}^2 I_{il}^2 \cos\left(\varphi_{ik} - \varphi_{il}\right) \right] \qquad (67)$$

Depending on powers, relation (67) becomes:

$$D_{all}^2 = \sum_{i=1}^{3} \left( S_i^2 - P_{nosym,i}^2 - Q_{nosym,i}^2 \right) \qquad (68)$$

Using relations (59) and (60), relation (68) becomes:

$$D_{all}^2 = \sum_{i=1}^{3} \left[ \left[ trace\left( \hat{\underline{U}}_i \hat{\underline{U}}_i^* \right) \right] \left[ trace\left( \hat{\underline{I}}_i \hat{\underline{I}}_i^* \right) \right] - \left( trace \hat{P}_i \right)^2 - \left( trace \hat{Q}_i \right)^2 \right] \qquad (69)$$

### 3.3.5. The non-symmetry deforming electric power

The non-symmetry deforming electric power is a concept introduced in paper [5] which represents, from a physical point of view the interaction between different harmonics that belong to different phases. The mathematical expression is:

$$D_{nosym}^2 = S^2 - P_{all}^2 - Q_{all}^2 - P_{nosym.all}^2 - D_{all}^2 \qquad (70)$$

Each of the terms from right is known from relations (61), (59), (60), (66) and (69).

## 4. Conclusions

In this paper, it is presented how the theory of the complex, linear operators for one and three-phased electric circuits can be applied when working in a non-sinusoidal periodic regime, without affecting the formulas from the sinusoidal periodic regime. For this purpose, there have been deduced:

In this paper, there have been introduced diagonal, infinite-dimensioned matrixes as complex linear operators with which there have been defined the following quantities:

- the operational form of the electric voltage $\hat{\underline{U}}$ and of the electric current $\hat{\underline{I}}$ relation (7)
- the operational form of the electric impedance $\hat{\underline{Z}}$ relation (8)
- the operational form of the harmonics $\hat{k}$ relation (12)
- the operational form of the effective values $|\hat{\underline{U}}|$ and $|\hat{\underline{I}}|$, relation (17)
- the operational form of the phase $\hat{\varphi}$, relation (22)
- the operational form of the an electric quantity in real time domain $\hat{i}(t)$, relation (26)
- the operational form of the complex power $\hat{\underline{S}}$, relation (28)
- the operational form of the active electric power $\hat{P}$ and of the reactive electric power $\hat{Q}$ relations (30)
- the operational form of the total active electric power $\mathbf{P_{all}}$ and of the total reactive power $\mathbf{Q_{all}}$, relation (31)
- the operational form of the apparent electric power, $\mathbf{S}$, relation (33)
- the operational form of the deforming power $\mathbf{D}$, relation (34)
- the operational form of the voltages on lines $\hat{\underline{U}}_{ij}$, relation (40)
- the operational form of the neutral point displacement voltage $\Delta\hat{\underline{U}}$, relation (43)
- the operational form of the rotation operators $\hat{1}, \hat{a}, \hat{a}^2$, relations (47), (48), (49)

- the operational form of the symmetrical components $\hat{\underline{U}}_h, \hat{\underline{U}}_d, \hat{\underline{U}}_i$, relation (50)
- the operational form of the impedances $\hat{\underline{Z}}_h, \hat{\underline{Z}}_d, \hat{\underline{Z}}_i$, relation (51)
- the operational form of the currents $\hat{\underline{I}}_h, \hat{\underline{I}}_d, \hat{\underline{I}}_i$, relation (54)
- the operational form of the currents on phase $\hat{\underline{I}}_1, \hat{\underline{I}}_2, \hat{\underline{I}}_3$, relation (55)
- the operational form of the apparent electric power $\hat{\underline{S}}_i$, relation (56)
- the operational form of the active electric power $\hat{P}_i$, relation (57)
- the operational form of the reactive electric power $\hat{Q}_i$, relation (58)
- the operational form of the total active power, $P_{all}$, relation (59)
- the operational form of the total reactive electric power $Q_{all}$, relation (60)
- the operational form of the apparent electric power, relation (62)
- the operational form of the non-symmetry electric power $P_{nosym,all}^2$, relation (66)
- the operational form of the deforming power $D_{all}^2$, relation (68)
- the operational form of the non-symmetry deforming electric power $D_{nosym}^2$, relation (70).